# Genomics models in radiotherapy:
# From mechanistic to machine learning

John Kang[1], James Coates[2], Robert Strawderman[3], Barry Rosenstein[4], Sarah Kerns[1]

[1]Department of Radiation Oncology, University of Rochester Medical Center, Rochester, NY, United States

[2]CRUK/MRC Oxford Institute for Radiation Oncology, University of Oxford, Oxford, United Kingdom

[3]Department of Biostatistics and Computational Biology, University of Rochester, Rochester, NY, United States

[4]Departments of Radiation Oncology and Genetics & Genomic Sciences, Icahn Institute for Data Science and Genomic Technology, Icahn School of Medicine at Mount Sinai, New York, NY, United States

# Abstract

Machine learning provides a broad framework for addressing high-dimensional prediction problems in classification and regression. While machine learning is often applied for imaging problems in medical physics, there are many efforts to apply these principles to biological data towards questions of radiation biology. Here, we provide a review of radiogenomics modeling frameworks and efforts towards genomically-guided radiotherapy. We first discuss medical oncology efforts to develop precision biomarkers. We next discuss similar efforts to create clinical assays for normal tissue or tumor radiosensitivity. We then discuss modeling frameworks for radiosensitivity and the evolution of machine learning to create predictive models for radiogenomics.

# Table of Contents

# 1  Introduction

In radiotherapy, radiogenomics is the study of the totality of genomic variation that affect the response of normal and tumor tissue to radiation.[1–3] There is significant interest in incorporating radiogenomics to guide personalize radiotherapy treatments.[4,5] Machine learning (ML) rose as a merger of computer science and statistics that provides a computationally-scalable framework for addressing high-dimensional classification and regression for prediction problems. ML is widely used in genomics for various tasks[6–8] and there are similar efforts to apply ML to high-dimensional radiogenomics datasets to create predictive models.[9] Like many fields in medicine, radiation oncology has taken significant interest in applying ML for several aspects of its practice.[10–14] Notable cooperative efforts in the United States include the American Association of Physicists in Medicine (AAPM) Practical Big Data Workshop (annually since 2017)[15] and a recently concluded National Cancer Institute (NCI) workshop on Artificial Intelligence in Radiation Oncology in April 2019.

In this review article, we begin with a background of genomically-guided therapies in medical oncology. We then contrast this with radiogenomics, beginning with efforts to predict normal tissue radiosensitivity before moving to the prediction of tumor radiosensitivity. Next, we discuss various frameworks for modeling radiosensitivity and how radiogenomics can augment these models. We then discuss the role and application of ML methods for specific radiogenomics applications to predict normal tissue and/or tumor radiosensitivity.

# 2  Precision oncology

While its precise definition is evolving, "precision oncology" generally is described as using genetic information to find targets for systemic agents.[16] Early examples of precision oncology focused on single mutations, such as imatinib to target *BCR-ABL* translocation in chronic myelogenous leukemia or trastuzumab to target *HER2-neu* oncogene amplification in breast cancer. In the last half decade, targeted therapies for specific tyrosine kinase mutations in lung cancer and melanoma have become standard of care.

## 2.1 NCI precision medicine trials

In March 2017, the US Food and Drug Administration (FDA) granted a tissue-agnostic "blanket approval" to the PD-1 inhibitor pembrolizumab for any metastatic or unresectable solid tumor with high microsatellite instability or deficiency in mismatch repair.[17] The rationale was the improved response rate and progression-free survival seen in colorectal tumors with genomic instability due to higher number of tumor neoantigens and lymphocytic infiltrates.[18] This approval marked the first time the FDA approved a drug for specific mutations independent of tissue type.

We can expect this pattern will only increase as the NCI is funding several "precision medicine" trials that are weighing mutational burden to guide therapy over the traditional method of being tissue-specific.[19] NCI-MATCH (for specific gene mutations) and NCI-MPACT (for mutations in specific pathways) are taking this tissue-agnostic, mutation-specific approach to targeted therapy.[20] The NCI-COG Pediatric MATCH is a similar to NCI-MATCH except patients will also have germline sequencing to determine whether the tumor genetic variants were inherited or not.

## 2.2 Biomarkers for decision making

In oncology, prognostic biomarkers provide treatment-agnostic information about a patient's outcome whereas predictive biomarkers provide information about whether a patient will benefit from a specific treatment.[21] The key point in the NCI-funded precision oncology trials is that regardless whether a biomarker is prognostic or predictive,[22] the result should alter the management of a patient. Even though prognostic biomarkers do not predict for benefit from a specific therapy, they can still change management in the right clinical context. In prostate cancer, serum prostate specific antigen (PSA) and pathologic Gleason grade are well-established prognostic biomarkers.[23] However, given that favorable PSA and Gleason grade are associated with a 99.9% 15-year survival from prostate cancer without upfront treatment,[24] the standard of care is to recommend against upfront radical treatment in patients in this very low risk category to avoid the toxicity of local treatment.[25] Similarly in gene panels, breast cancer patients who met criteria for adjuvant chemotherapy but had a low Oncotype DX[26] Recurrence Score® ≤11 were prospectively given hormone therapy alone and observed to have 5-year recurrence rate of only 1-2%.[27] Even in the absence of randomized evidence for this low risk category, the standard of care is to avoid the toxicity of chemotherapy in this low risk group.

In general, prognostic models that were not trained on an untreated cohort have limited ability to alter management as it can be difficult to distinguish causal effect of treatment vs. favorable biology of a treated cohort. For example, Speers et al. created a radiosensitivity signature for breast cancer local recurrence by training and validating on two cohorts treated with adjuvant radiation.[28] The authors note, however, that their signature is unable to differentiate between a radioresistant cancer and a cancer with aggressive biology. Similarly, this signature would be unable to differentiate between a radiosensitive cancer and a cancer with indolent biology.

Further confusion occurs when a biomarker appears to have both prognostic and predictive value. While temozolomide chemotherapy has shown to significantly increase overall survival in O-6-methylguanine-DNA methyltransferase (MGMT) gene promoter-methylated glioblastoma, this benefit is significantly attenuated (if it exists at all) in non-MGMT promoter-methylated glioblastoma.[29,30] However, the methylation itself seemed to have prognostic significance independent of temozolomide, with the recent NOA-09 trial of only MGMT promoter-methylated patients showing unprecedented survival of 31 months in the standard arm.[31,32]

While there can be vigorous debate about whether a biomarker is predictive or prognostic,[33–36] the focus should remain on the potential for a biomarker to alter clinical management.

# 3 Normal tissue radiosensitivity

Radiation oncology generally aims for 5-10% clinically-significant toxicity to organs at risk in clinical trials. As a result, the patients with the most radiosensitive normal tissue ultimately determine the limit for the maximum dosage for all patients.[4] It is no coincidence that the adjacent organ-at-risk dose tolerance and the recommended prescription dose mirror each other for most solid tumors. In contrast, medical oncology trialists will often accept in excess of 60% grade 3+ toxicities for advanced or metastatic disease.[37–39] These studies suggest that patients with more advanced disease are more willing to accept higher toxicity for an improved probability of survival, which is in agreement with a focus on better prediction of individualized toxicity.

## 3.1 Improving treatment planning

Normal tissue radiotherapy response depends on several factors. Ever since the early days of isoeffect curves, we have known that different tissue types have different radiosensitivities. To address this, organ-specific guidelines were published in the QUANTEC (Quantitative Analysis of Normal Tissue Effects in the Clinic) special edition of the Red Journal in 2010.[40] However, much progress has been made in the 9 intervening years. QUANTEC was published during a transition period from 3D conformal radiation to intensity-modulated radiation therapy (IMRT), yet QUANTEC dose tolerances have commonly carried over for IMRT planning. Given the improved conformality of moderate-high dose region in IMRT, these constraints likely represent overly conservative measures for most organ sites. Hypofractionation—both moderate and stereotactic-level—was also not addressed but has since become standard of care for several of the most commonly treated organ sites. ASTRO guidelines for hypofractionated or stereotactic treatment have published within past 2 years for lung,[41] prostate,[42] and breast cancer.[43] Since QUANTEC, there have been similar efforts by PENTEC[44] (Pediatric Normal Tissue Effects in the Clinic) and the AAPM working group HyTEC[45] (Hypofractionation Treatment Effects in the Clinic) to determine normal tissue effects for pediatric patients and hypofractionated regimens, respectively.

## 3.2 Genomic basis for radiosensitivity

While these advances by QUANTEC, PENTEC and HyTEC hold significant promise, they focus primarily on dose constraints and largely omit non-dosimetric clinical factors which can affect organ function following radiotherapy, such as diabetes, vasculopathy, hypertension or smoking.[46] Efforts from the 1980-1990s studying skin toxicity in breast cancer suggested a patient's toxicity could be tissue-specific,[47] temporally-specific[48] (i.e., acute did not necessarily predict for late toxicity) and dominated by a genetic as opposed to a dosimetric component.[49] These works provided early evidence of genetic predisposition to radiosensitivity and suggested that there were not only rare-yet-highly-penetrant radiosensitivity genes (i.e. *ATM*), but also more subtle genetic biomarkers that do not manifest until directly insulted with radiotherapy.

### 3.2.1 SNP analysis: candidate gene

With the advent of genome-wide genotyping and next-generation sequencing in the 2000s, cost-effective and high-throughput efforts to investigate the genetic component of radiotherapy toxicity became possible. Analysis with single nucleotide polymorphisms (SNPs), which are single nucleotide variations between individuals, was—and still is—a popular method due to its low cost and ability to capture the majority of common (allele frequency ≥ 5%) SNPs across the genome. Most of the early radiogenomics studies used a candidate gene approach where SNPs were pre-selected for their proximity to genes known to be involved in cellular radiosensitivity. Early candidate gene studies were often small, underpowered, and performed with a poor understanding of important statistical issues, namely, false discovery due to uncorrected multiple hypothesis testing.[50] Unsurprisingly, validation studies with independent cohorts would fail to replicate these associations and suggest publication bias.[51,52]

In response, the international Radiogenomics Consortium (RgC; epi.grants.cancer.gov/radiogenomics/) was formed in 2009 to expand knowledge of the genetic basis for radiosensitivity differences in patients and to develop assays to help predict susceptibility for the development of adverse effects from radiotherapy.[53,54] The RgC aims to foster international collaborative projects, share biospecimens and data, develop guidelines to improve the standardization of radiogenomics research, and provide a forum and framework for discussion, development and pursuit of new research directions. In recognition of the dearth of new junior investigators in the field, the RgC also supports the development of early career investigators and facilitates mentoring relationships across academic institutions. While achieving large sample sizes for cohorts with detailed radiation dosimetry, toxicity, and SNP data is an ongoing challenge, the RgC has been key in increasing study sizes and, correspondingly, statistical power. Newer candidate gene analyses have generally validated findings on independent cohorts. RgC members have found and validated SNPs associated with late skin, fibrotic, and overall toxicity after breast cancer radiotherapy upstream to inflammatory cytokine *TNFA*[55] (rs1800629, rs2857595) and within base-excision repair gene *XRCC1* (rs2682585).[56] A large meta-analysis validated the association of rs1801516 in DNA double-strand break (DSB) repair gene *ATM* with several acute and late toxicities in breast and prostate cancer, thus mimicking a milder phenotype of ataxia-telangiectasia syndrome.[57] The Liao group has identified and validated SNPs to predict for Grade ≥3 toxicity after lung cancer radiotherapy. These include SNPs in heat shock protein *HSPB1* associated with pneumonitis[58] and esophagitis,[59] and rs1800469 in profibrotic cytokine *TGFB1* associated with esophagitis.[60] The role of the rs1800469 SNP remains unclear, and may be tissue-specific. RgC members performed an individual-patient meta-analysis of 2782 patients from 11 cohorts to assess for association of rs1800469 with late fibrosis or late overall toxicity (using a harmonized average score) in breast radiotherapy and found no association with multivariable odds ratio 0.98 (95% confidence interval 0.85-1.11).[61] However, Grossberg et al. report a positive association following breast radiotherapy with odds ratio 4.47 (95% confidence interval 1.22-15.99) as assessed with the Late Effects Normal Tissue/Subjective, Objective, Medical Management, Analytic scale.[62]

### 3.2.2 SNP analysis: genome wide association studies

An alternative to candidate gene studies is genome-wide association analysis (GWAS). Contrary to the targeted candidate gene approach, GWAS casts a wide net to sieve out moderate levels of signal in noisy genomic data.[52] The threshold for statistical significance is more stringent to account for the increased number of independent hypotheses performed in GWAS (one per SNP) and thus larger sample sizes are typically required to find statistically-informative associations. Recent GWAS, including meta-analysis, by RgC members have identified several novel SNPs associated with genes which were not previously linked to radiotherapy toxicity.[63] These genes include *TANC1*, which encodes a repair protein for muscle damage,[64] and *KDM3B* and *DNAH5*, which encode proteins expressed in urinary tract and were associated with increased urinary frequency and decreased urinary stream, respectively.[65]

### 3.2.3 Beyond SNPs

Copy number variation (CNV) or gene dosage is another genetic alteration that is estimated to account for 4.8-9.5% of the genome.[66] A CNV exists when a DNA segment 1 kb to several Mb in length is either deleted or duplicated compared to a reference genome.[67] The objective of CNV analysis is to determine how many copies of a gene regions deviates from two, representing a reference gene with one copy from each homologous chromosome. CNVs have been implicated in several diseases, but their role in radiotherapy toxicity is nascent. In a proof-of-concept study, Coates et al. incorporated *XRCC1* SNP and CNV features into classic models to improve performance.[68]

In a very large pre-clinical study, Yard et al. illustrated the rich space of CNVs, gene mutations, and gene/gene set expressions by exposing over 500 cell lines to radiation.[69] Radiosensitivity was enriched in gene sets associated with DNA damage response, cell cycle, chromatin organization, and RNA metabolism. In contrast, radioresistance was associated with cellular signaling, lipid metabolism and transport, stem-cell fate, cellular stress, and inflammation.

# 4 Tumor radiosensitivity

Tumor radiogenomics has focused on finding gene expression assays to determine radiosensitivity.[3,70] The general goal of these assays or models is to predict whether an individual benefit would be more served by radiotherapy compared to an alternative treatment.

Tumor genomics should be distinguished from germline genomics as the former is usually using genetic information from tumor tissue whereas the latter is using DNA taken from normal tissues, typically blood or saliva (**Table 1**). There are exceptions as described below with *MRE11A* and later with multi-omics models by the El Naqa group.[71]

**Table 1**: Examples of tumor gene expression assays/models and consortia for germline polygenetic assays

| Poly-gene expression assays (tumor behavior) | Polygenic germline genomics (normal tissue risk) |
| --- | --- |
| **General oncology** | **Cancer development risk** |
| Breast: Oncotype DX®, MammaPrint® Prostate: Decipher®, Prolaris® | Consortiums for prostate cancer (PRACTICAL), breast cancer (BCAC), ovarian cancer (OCAC) |
| **Radiation oncology** | **Radiotherapy toxicity risk** |
| Prostate: post-operative radiation therapy outcomes score (PORTOS) Various tumors: genome adjusted radiation dose (GARD) | Radiogenomics Consortium |

## 4.1 Gene expression to predict radiotherapy benefit

Pre-treatment bladder cancer expression of DSB repair gene *MRE11* was found to be predictive of survival following radical radiotherapy but not cystectomy by the Kiltie group,[72] a result that was independently verified by Laurberg et al.[73] Interestingly, the SNP rs1805363 in *MRE11A* was also found predictive of survival in a gene-dosage effect in patients with bladder cancer treated by radical radiotherapy but not by cystectomy.[74] New work by the Kiltie group aimed to standardize and develop a clinical assay for MRE11 immunohistochemistry assay across several centers in the United Kingdom using transurethral resection of bladder (TURBT) samples from prior bladder-preservation trials.[75] However, they were unable to replicate prior results, which was attributed to problems in automatic staining and poor scoring reproducibility among histopathologists.

The Post-Operative Radiation Therapy Outcomes Score (PORTOS) is a 24-gene predictive assay that was retrospectively trained on post-prostatectomy prostate cancer samples from patients, some of who received postoperative radiotherapy.[76] The authors used gene ontology to find gene sets enriched for DNA damage and radiation response. Using a final pool of 1800 genes, a univariable screen was used for ranking, followed by ridge regression-penalized Cox modeling of distant metastasis free survival to determine which patients would benefit from post-operative radiation therapy.

The radiosensitivity index (RSI) was developed at Moffitt Cancer Center to predict radiosensitivity in several tumor cell lines.[77] The RSI signature is a linear regression of the expression of 10 genes (*AR, cJun, STAT1, PKC, RelA, cABL, SUOMO1, CDK1, HDAC1, IRF1*) with the outcome variable being survival fraction at 2 Gy (SF2). The selected genes are implicated in pathways involved in DNA damage response, histone deacetylation, cell cycle, apoptosis and proliferation. In patients, RSI predicted for improved progression free survival in cohorts treated with adjuvant radiation but *not* in unirradiated controls in several tissue types, including breast cancer,[78] lung cancer,[79] and prostate cancer.[80] More recently, Scott et al. combined RSI with the linear quadratic model to create the Genome Adjusted Radiation Dose (GARD).[81] The goal of GARD is to unify a model of both radiobiologic and genomic variables to predict for radiation response, and provide a quantitative link from genomics to clinical dosing. There is discussion regarding whether GARD is truly prognostic or predictive as the radiation dosage component is a modifiable input.[33,34]

Another class of molecules that has gained interest are micro RNAs (miRNA) which are short, stable sequences of RNA which act to inhibit expression of messenger RNA (mRNA).[82] Pretreatment serum levels have been found to correlate with radiotherapy response cancers of the lung[83] and head and neck.[84,85] miRNA upregulation can occur just hours after ionizing radiation.[86]

## 4.2 Prognostic or predictive radiation signatures

As there was no alternative treatment that patients could have received instead of radiation in the populations used to train RSI/GARD or PORTOS, it remains unclear whether these tests are acting as specific biomarkers of radiotherapy benefit or whether they serve as general markers for treatment sensitivity. In other words, perhaps a similar benefit or even increased benefit would be seen with systemic therapy or other therapy. Without comparison to alternative therapies, there will remain a question of whether radiation is the correct choice in a space increasingly crowded with targeted therapies, immunotherapy, and salvage surgery. As the MRE11 studies did train models on patients who received either radical radiotherapy or cystectomy, it would be correct to say that MRE11 expression does predict for radiotherapy benefit vs. surgery. However, it should also be cautioned that this predictive ability could be due to confounding variables or other latent factors, particularly as initial clinical validation was unsuccessful (discussed above[75]). Thus, while MRE11, PORTOS, and RSI/GARD remain to be prospectively or randomly validated, they are a promising harbinger of future radiogenomics assays to determine benefit from radiation.

Tumor sensitivity assays remain of active interest. In the newly published Children's Oncology Group ACNS 0121 phase 2 trial, improved event free survival (EFS) was seen in pediatric infratentorial ependymomas receiving chemoradiation who had focal copy number gain on chromosome 1q compared to those without 1q gain (83% vs. 47%, p=.0013).[87] As there was no cohort with focal 1q copy number gain that did not receive chemoradiation therapy, we do not know whether this mutation is predictive of chemoradiation response or general prognostic indicator.

# 5 Radiogenomic modeling: mechanistic, data-driven and machine learning methods

The following section highlights methods and examples of different frameworks for radiogenomics modeling that will be discussed in this section (**Figure 1**). We first discuss augmentation strategies for "classical" (dose- and volume-based) outcomes models to allow the integration clinical and biological data ("augmentation"). We then discuss data-driven modification of these models as well as data-driven models that use machine learning.

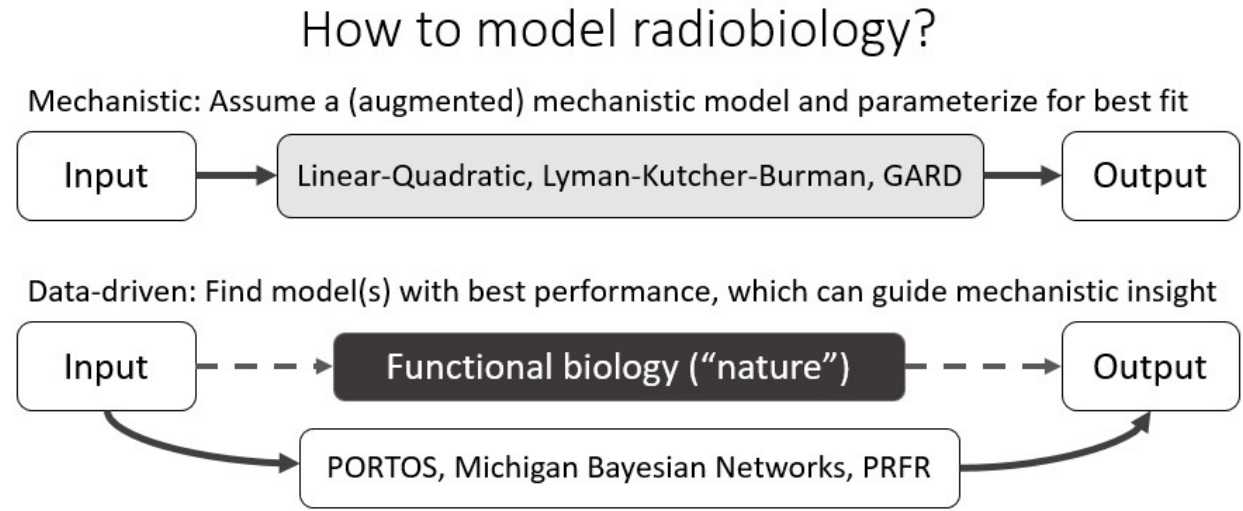

**Figure 1**: Schematic comparing modeling frameworks for mechanistic vs. data-driven models with select examples discussed in this review. GARD: Genome-Adjusted Radiation Dose[81]; PORTOS: Post-Operative Radiation Therapy Outcomes Score[76]; PRFR: preconditioned random forest regression[88]

## 5.1 Classical models

Normal tissue complication probability (NTCP) models attempt to predict the risk of specific normal tissue side effects induced by radiotherapy. Examples include radiation fibrosis in the lungs or erectile dysfunction after prostate radiotherapy. These toxicities reflect both cell death as well as non-cell death related tissue injury, including inflammatory reactions and late effects. Until recently, such predictive models have relied almost exclusively on semi-mechanistic interpretations of dose and volume data to understand the effects of ionizing radiation.[89,90] In this paper, we refer to "mechanistic" models of NTCP to describe these simplified, semi-mechanistic models, such as the Lyman-Kutcher-Burman model.[91] It is now apparent that more complex genetic and biological factors can play a significant role in determining and modulating tissue radiosensitivity, often in non-linear ways.[92] Pragmatically, NTCP models based on dosimetry-alone risk lacking robustness across cohorts potentially limiting their usefulness clinically.

Biological risk factors integrated into classical outcome models generate "augmented" (or "mixed") models. This approach can also be extended to clinical risk factors as well. The impact of integrating a validated biological risk factor is visualized in **Figure 2a** in the change of NTCP curve slope and absolute value according to different generalized equivalent uniform doses (gEUDs). A corresponding shift is showcased in **Figure 2b** for the same theoretical, dichotomous biological risk factor (present or absent) but is plotted with separate dose and volume components in three-dimensional space. These figures use mock input data but show changes similar in magnitude to those in literature.[92–94]

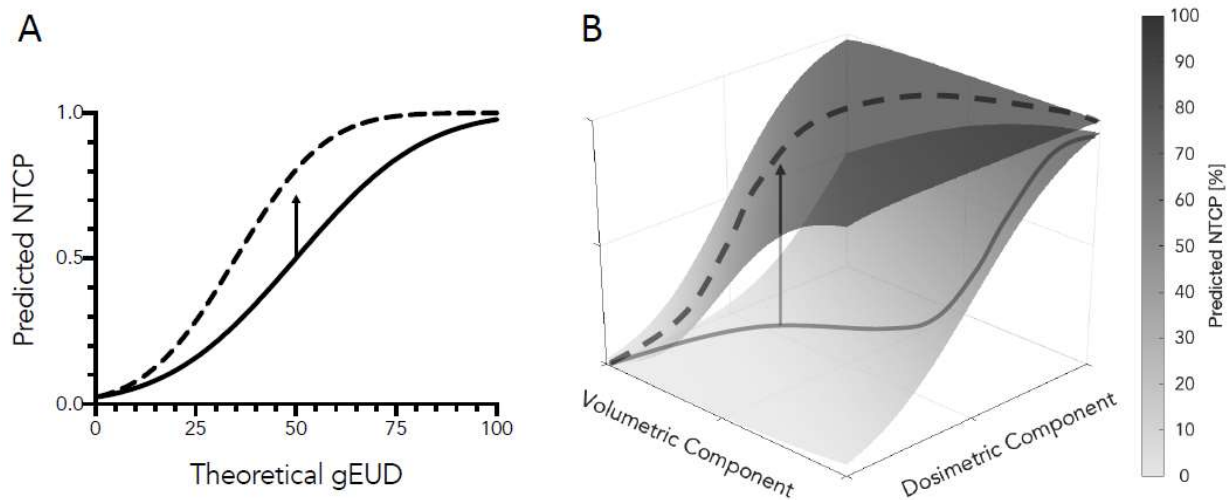

**Figure 2**: Theoretical impact of integrating validated biological risk factors on predicted NTCP curves (a) and surfaces (b). Solid curves indicate theoretical NTCP values derived from dose and volume metric-based models. Dashed curves reflect a dosimetric model that has been augmented with a dichotomous biological risk factor (radiogenomic). In (a), NTCP values were generated using the Lyman-Kutcher-Burman (LKB) model with mock generalized equivalent uniform dose (gEUD) values yielding NTCP values with or without a risk factor (as indicated). Surfaces in (b) were similarly generated with mock values for dose and volume parameters making separate gEUD-based planes.

In order to generate NTCP curves, there exists two overarching modeling approaches: data-driven and mechanistic. Data-driven approaches use statistical techniques to identify trends and correlations in data while mechanistic approaches[95,96] (also known as analytical techniques) rely on certain simplifying assumptions to reflect experimental observations. The methodology for augmenting a model can vary widely depending on the chosen strategy and is therefore equally important to consider. In the below sections, we discuss types of biological factors used in augmenting models and associated strategies with relevant examples.

## 5.1.1  Integration of biological variables

SNPs reflect single changes in nucleotide sequences and their incorporation into NTCP models has become more common recently although most findings remain to be validated on large cohorts. If found in coding regions, SNPs may induce conformation or enzymatic changes to transcripts or proteins. For NTCP modeling, these changes could affect DNA repair capabilities, signal transduction, or other elements of the DNA damage response (DDR) cascade.[97]  In the case of SNPs that do not reside in protein coding regions, mechanistic influence on outcomes is less clear but could theoretically relate to transcript splicing, binding, or degradation.[98] As discussed above, SNPs of interest were historically sought using candidate gene approaches and most reports related to DDR pathways.[2,99]

Previously, our group demonstrated that more substantial changes to the genome could be useful for outcome modelling in addition to SNPs for predicting late effects in prostate cancer radiotherapy.[68] Resulting radiogenomic models incorporating SNPs, CNVs, and dosimetry led to increased cross-validated predictive power when compared to dosimetric- or genetic-only models. The CNV of DNA repair gene XRCC1 was found to be particularly useful for predicting severe rectal bleeding (Grade ≥3) or erectile dysfunction (Grade ≥1) when combined with dosimetry. Though improvement in cross-validated prediction performance was found for both data-driven and mechanistic modeling strategies, performance was higher in the case of data-driven models compared to mechanistic ones. Given the limited sample size (n=54), the worse performance of the mechanistic model may reflect incorrect assumptions introducing a bias, which can be compensated for in the data-driven model. Thus, the relative difference in performance observed in the study may be dependent on the modeling frameworks employed.

In the case of augmenting existing mechanistic models, it is important to consider how variables such as the binary presence of genetic variants are incorporated mathematically to avoid nullifying model output in the absence of the risk factor; this principle applies similarly to clinical risk factors. The use of dose-modifying factors (DMFs) maps existing parameters into an exponential function to avoid misbehaved outputs (e.g., when it goes to zero or ∞).[68,93,100] In contrast, data-driven approaches are readily able to integrate dichotomous and continuous data.

### 5.1.2 Modulating dose in classical models

The most frequently employed dosimetry-based NTCP model in use is the Lyman-Kutcher-Burman (LKB) model (**Eq. 1a-c**):[101]

$$NTCP = \Phi\left(t\right) \tag{1a}$$

where $\Phi(t)$ is the normal cumulative distribution function:

$$\Phi\left(t\right) = \frac{1}{\sqrt{2\pi}} \int_{-\infty}^{t} e^{\left(\frac{-u^2}{2}\right)} du \tag{1b}$$

consisting of:

$$t = \frac{EUD - TD_{50}}{m \cdot TD_{50}} \quad \text{and} \quad TD_{50}(V) = \frac{TD_{50}(1)}{V^n} \tag{1c}$$

where $TD_{50}(1)$ is the dose at which NTCP is 50% for an endpoint, m the slope, and $TD_{50}(V)$ is the partial volume tolerance dose, and $n$ is the tissue-specific volume exponent.

Formulaically, the LKB model is based on a sigmoidal error function and stratifies patient risk according to how much larger or smaller the gEUD is relative to the $TD_{50}$. The gEUD is a three-dimensional DVH reduction technique and so spatial information on dose distribution is not fully retained; the dimensionality of the problem is significantly reduced by sacrificing details of the spatial dose distribution (see **Section 5.4.3** for relevant discussion on feature transformation). Consequently, the LKB model and its variants (described below) best describe toxicities related to uniform doses of all or part of an organ. Depending on whether the canonical LKB model is employed and the treatment conditions or endpoint considered, the LKB model is sometimes referred to as semi-mechanistic due its relatively simplistic formulation derived from the normal cumulative distribution function. The modified LKB model[102] (**Eq. 2a-b**) allows for the inclusion of biological risk factors whether they are dichotomous or not:

$$NTCP = \frac{1}{\sqrt{2\pi}} \int_{-\infty}^{t} e^{\frac{-u^2}{2}} \, du$$

$$t = \frac{D_{eff} - TD50 \cdot DMF_1 \cdot DMF_2 \cdot ... \cdot DMF_k}{m \cdot TD50 \cdot DMF_1 \cdot DMF_2 \cdot ... \cdot DMF_k}$$

(2a)

where each DMF is defined by:

$$DMF_k = e^{\delta_k \cdot R_k}$$

(2b)

wherein $\delta_k$ is the risk-factor weight for risk-factor $k$, and $R_k$ indicates the presence ($R_k$=1) or absence ($R_k$=0) of the risk-factor for a given patient.

The GARD/RSI model discussed above is a practical example of augmenting the linear-quadratic model of cell kill by adding gene-expression derived parameters to create a patient-specific radiosensitivity assay.[81]

## 5.2 Data-driven methods: statistical models

Data-driven models (also known as phenomenological models) are built on statistical techniques that are empirical and frequentist.[11] Classical models employ a more rigid mathematical basis (typically around a mechanistic basis) around which goodness-of-fit is sought, whereas data-driven statistical models have less restrictions and assumptions. The robustness of data-driven models relates to how often observations (data or variables) can be related to an outcome according to certain relationships (equations). As such, data-driven NTCP models may be able to identify models that perform better than mechanistic counterparts, given sufficient data, or that depart significantly from standard therapeutic strategies, such as charged particle therapy.[103]

Data-driven NTCP models are most often regression-based and sigmoidal in shape. The reasons for this are that regression-based techniques can be trained in a practical amount of time such that prospective clinical use is feasible. Secondly, the experimentally observed sigmoidal relationship seen with ionizing radiation and cell death can be reflected by regression-based models by using sigmoidal link functions.

Let $\pi(x_i)$ be the desired probability considered as a function of the input variable(s) $x_i$. Then, the most commonly used link functions are the inverse of the normal cumulative distribution function, or probit (**Eq. 3**), and the logit function (**Eq. 4**):

$$\Phi^{-1}\big(\pi(x_i)\big) = g(x_i)$$

(3)

$$\log\left(\frac{\pi(x_i)}{1-\pi(x_i)}\right) = g(x_i) \tag{4}$$

where $g(x_i)$ is the generalized linear formulation of the input variable(s), $x_i$:

$$g(x_i) = \beta_o + \sum_{j=1}^{s} \beta_j X_{ij}, i = 1,...,n, j = 1,...,s \tag{5}$$

having β as coefficients that are identified via optimization, e.g. maximum likelihood estimation.

Increased generalizability for data-driven models may also originate from frameworks having to first choose whether to include a variable or not (model order estimation) prior to fitting coefficients.[104] In contrast, mechanistic modelling approaches require that some or all parameters be included and therefore only be modulated, which may limit generalizability due to forcing features into the model.

The PORTOS assay discussed above is a practical example of augmenting Cox regression survival analysis with gene-expression derived tumor radiosensitivity.[76]

## 5.3 Data-driven methods: interpretable machine learning

Machine learning (ML) evolved from a recent cross-pollination between computer science and statistics. As owed to their computer science origin, most ML models are "data-driven" techniques as they do not attempt to model simplified mechanisms[105] (as is done in the mechanistic/analytical approach) and instead statistically translate inputs into probabilities that appear sigmoidal, though the specific link function is a user-set hyperparameter. Due to the focus on using data to make predictions but not in inferring the processes behind data generation, many ML methods are rightly described as having poor interpretability. Improving interpretability is of particular value in clinical decision support models[106–109] and is a high priority research area in ML, being the subject of a Neural Information Processing Systems symposium in 2017.[110]

Some data-driven models in ML attempt to provide a more interpretable context. The progression of Bayesian network approaches by the El Naqa group provides a case study of more interpretable, multi-objective modeling in radiogenomics, as described below.

### 5.3.1 Bayesian networks

Bayesian networks (BN) are directed graphical models with probability distributions.[111] The graph is represented by nodes (vertices) connected by links (edges) that represent probabilistic relationships between the variables. The graph then captures joint distributions between subsets of the variables. In fact, relationship between nodes in a BN can be deconstructed into a series of joint probabilities. BNs are subject to an important restriction of being acyclic: there must not be a closed path from a node to another node along links following the direction of the arrows. Despite their name, they can be used for either frequentist or Bayesian inference.[112]

Like decision trees, BNs are graph-based methods that are reasonably interpretable and can incorporate variables from different domains without extensive pre-processing. A disadvantage is that continuous variables are typically discretized as many BN learning algorithms cannot efficiently handle continuous variables and common software packages do not support continuous variables.[113] In radiogenomics, BN nodes represent genes or proteins while the links between them represent probabilistic similarities or interactions.[70] Due to these advantages, BNs have been used to incorporate both clinical and genomic variables to predict local control and/or toxicity.

### 5.3.2   Single objective prediction: local control or toxicity

Oh et al. used BN on two separate patient cohorts of locally-advanced non-small cell lung cancer (NSCLC) to predict local control (LC).[71] The first cohort of 56 patients included clinical and dosimetric data. The second cohort of 18 patients also had serum markers drawn pre- and during-treatment: transforming growth factor β, interleukin 6, angiotensin converting enzyme, and osteopontin. Despite the small numbers, better performance was noted when clinical, dosimetric, and molecular data were all used.

This BN framework was also applied to 54 locally advanced NSCLC patients to predict Common Terminology Criteria for Adverse Events (CTCAE) v3.0 Grade 2+ radiation pneumonitis (RP2). Initially, a limited set of pre- and during-treatment molecular biomarkers was used along with clinical/dosimetric features (i.e., lung V20)[114] but this was subsequently expanded to a heterogeneous set of 200 features, which included SNPs, cytokines and micro RNA (miRNA) biomarkers.[115] Feature selection was performed using a stepwise wrapper method[116] to find conditionally independent subsets (Markov blankets[117]). Predictive performance for these studies reached area under the receiver operating curve (AUROC) >0.80.

### 5.3.3   Multi-objective prediction: local control and toxicity

Prior radiogenomics studies have generally focused separately on either tumor control or toxicity. More recently, Luo et al. demonstrated a multi-objective approach using BN to simultaneously model LC and RP2 in a cohort of 118 NSCLC patients treated with radiotherapy.[118] A total of 297 clinical, dosimetric, radiomic, and genomic features were initially considered. The genomic features were pre-selected from literature review and included micro RNA (miRNA), SNPs, and cytokine levels in serum. The cytokine and the metabolic radiomic features had values both pre- and during-treatment. By first training individual BNs for LC and RP2, the authors were able consider a smaller feature set of about 50 (i.e., only features important for either LC or RP2) to optimize a unified multi-objective BN that simultaneously modeled LC and RP2. This process was repeated using pre-treatment or during-treatment variables. While the validation set was heavily skewed with only 3/50 patients having RP2, this work does provide a conceptual framework that accounts for both LC and toxicity.

### 5.3.4   Challenges of modeling therapeutic ratio

The therapeutic ratio (therapeutic index) refers to the ratio of tumor response for a fixed level of normal tissue damage.[119] Like many tradeoffs in statistics and ML—bias/variance, sensitivity/specificity—the multi-objective tradeoff between tumor control and toxicity in modern treatment planning follows[5] the principles of Pareto optimality or efficiency where one outcome cannot be improved without worsening another.[120]

While the work by Luo et al. demonstrates the feasibility of a multi-objective ML model that optimizes a joint AUROC, it does not allow one to give relative weight to LC vs. toxicity. There are several complicated issues to consider. First, LC and toxicity are not independent events as local recurrence in lung cancer after first-line therapy generally leads to additional salvage treatments, which could include additional radiation, surgery and/or systemic therapy. The toxicity profile is now magnified with no guarantee of a linear increase in toxicity. Secondly, RP2 may be an overly conservative endpoint as CTCAE v3.0 Grade 2 pneumonitis is defined as "symptomatic, not interfering with ADL." This issue relates to the differing weights that patients and physicians place on LC and toxicity. Patients may value LC more than toxicity due to fear of recurrence[121,122] whereas physicians may weigh toxicity more than LC.[123] Whereas a tumor recurrence may be attributed to aggressive biology, physicians can be more certain of therapy-induced toxicity, which can induce recall bias of patients who had severe toxicity. Lastly, it can be very difficult to assess adverse events. Clinicians tend to underreport subjective treatment toxicities compared to patients, which is the impetus behind patient-reported outcomes.[124,125] In the setting of a multi-objective model, being able to precisely define toxicity becomes of upmost importance due to the direct relationship with tumor control. Wider adoption of patient-reported outcomes could partially alleviate this issue. The addition of transcriptomic or proteomic information from the tumor tissue itself (instead of circulating cytokines) would also likely improve the assay.

In the future, we will likely see tumor expression panels such as PORTOS and GARD/RSI combined with germline toxicity panels to determine multi-objective optimization for therapy aggressiveness. Such radiogenomic approaches that can increase the therapeutic ratio are sorely needed.

## 5.4 Data-driven methods: high dimensionality machine learning

While many of the classic ideas and methods in machine learning (ML) were developed several decades ago, the exponential increase in computing power[126] has recently allowed high-dimensional ML models to be trained in a reasonable timeframe. However, there is often interest in only using certain features of the dataset. Libbrecht and Noble describe three reasons for feature selection in machine learning for genomics: (1) identify a subset of features that retains good predictive performance, (2) understand biology through the principle of parsimony, and (3) improve performance by removing variables that are contributing noise.[7] The authors note that it is usually very difficult to perform all three simultaneously. Feature selection in genomics is difficult not only due to the large number of features (i.e. SNPs, CNVs, etc) involved, but also due to the complex biology represented. This complexity can result in low "signal to noise" that can make predictive models difficult to train.

High-dimensional feature selection is an active area of research in genomics.[127] More generally, it has long been an active area of research among statisticians, particularly so for regression problems. Arguably, these efforts began with the introduction of forward variable selection,[128] where it has since been established that forward and related methods of stepwise regression (i.e., wrapper methods) are fraught with numerous inferential challenges.[129] Modern approaches to variable (i.e., feature) selection in regression and classification vary in scope and complexity, including approaches that utilize univariate screening (or filtering) procedures,[130,131] penalized regression or classification methods (e.g., lasso,[132] elastic net[133]), and combinations thereof.[134] The statistical properties of these newer methods are necessarily studied under certain sparsity assumptions, where (i) the number of available predictor variables (e.g., SNPs) can vastly outstrip the number of independent replicates (e.g., subjects) but (ii) the number of important variables is much smaller. Despite the many important advances in both estimation and inference in the presence of variable selection that has been made since the introduction of the lasso, numerous unresolved challenges remain and research into these problems continues.

Consequently, feature selection in genomics relies on custom combinations of filters, wrappers, embedded methods, and dimensionality-reduction. These methods have their own advantages and disadvantages that will be further discussed (**Table 2**). The optimal process of sequencing and combining feature selection methods with domain expertise has not been well established and is more of an "art."

**Table 2**: Feature selection methods discussed in Section 5.4 with examples, advantages and disadvantages. PCA: principal component analysis.

| Feature selection method | Advantage | Disadvantage |
|---|---|---|
| Filter (t-test, chi-squared) | fast; statistical inference | ignores interactions |
| Wrapper (stepwise selection) | multivariable interactions; widely implementable | slow; inferential challenges |
| Embedded (lasso, elastic net) | balances complexity with performance | model-specific; assumes sparsity; inferential challenges |
| Transformation (PCA, manifold learning) | partially preserves information; uses unlabeled data | unclear interpretability |

### 5.4.1  Filtering: pre-processing independent of model

In filtering, the features are pre-selected using a type of screen that is independent of the model choice. Filtering increases robustness to overfitting by reducing variance at the expense of increasing bias:[135] there is an assumption that the remaining variables are sufficient for explaining a model's behavior. There is usually also an assumption that correlation between variables can be ignored. Prior to formal analysis, GWAS studies use quality control filters which can consist of testing for SNPs out of Hardy-Weinberg equilibrium, with a high missing proportion, or with a low minor allele frequency.[136,137]

In high dimensional problems scenarios like GWAS, millions of hypothesis tests are performed. By random chance, there is a near-guarantee of finding effects that are falsely labeled as statistically significant if there is no correction to the pre-specified significance level $\alpha$.[138] Several techniques are commonly used in GWAS to correct for the probability of Type I errors, which is the risk of rejecting a true null hypothesis.[139] A Type I error can also be described as a false positive. Two important methods for controlling Type I error include the familywise error rate (probability of at least one Type I error) and the false discovery rate (the expected proportion of false discoveries).[140] One common familywise error rate method is Bonferroni correction, which involves dividing the $\alpha$ of a single hypothesis test (i.e., the threshold for obtaining a surprising result assuming no difference; often set at $\alpha = 0.05$) by the number of independent tests performed. For example, in a GWAS with approximately 1 million independent SNPs being evaluated individually, the corrected $\alpha$ would be set to $0.05 \div 10^6 = 5 \times 10^{-8}$. Use of Bonferroni correction to account for multiple hypothesis testing has become standard practice in genetic association studies, including those focused on radiotherapy outcomes.[65,141]

The main weakness of univariable filtering is that it does not account for interactions between features and thus is unable to explain non-linear behavior between features, such as epistatic interactions.[142] Epistatic interactions are difficult to find and interpret in genomics, as a statistical interaction does not necessarily imply a biological interacton.[127,143,144] In the most extreme scenario, a variable may have no marginal effect (i.e., no independent effect or main effect) and only be detected through an interaction when combined with another variable. To address this weakness in univariable filtering, multivariable methods such as the ReliefF family take an ensemble approach to rank variables.[145–147] Other options include incorporating feature selection into model building, as discussed below.

### 5.4.2 Wrapper and embedded feature selection

Wrapper methods and embedded feature selection methods both incorporate feature selection into the model building. Wrappers are model-agnostic methods that select among feature subsets by incorporating ("wrapping") the model being assessed into the feature selection process.[148] Examples include stepwise forward selection, backward elimination, and genetic algorithms.[135]

Embedded methods are model-specific and typically multi-objective in that they balance predictive performance with model complexity. An example is regularization, which adds a penalty parameter to a method's objective function to control for complexity. While superficially similar to information theory approaches in that both control for complexity, regularization is part of the internal model optimization process whereas information criterion approaches are used for model selection.[149] Genomics uses regularization extensively to decrease the polygenic complexity and decrease variance.[150] Common usages include linear regression regularization aimed to prevent the beta coefficients from growing too large (ridge regression[151]), to shrink the number of non-zero beta coefficients (lasso regression[132]), or both (elastic networks[133]).[152,153] A related two-step method called stability selection was developed for very high dimensional variable selection.[154] A subsampling step is first used to determine the amount of regularization and is then followed by a lasso step.

Due to their ability to consider epistasis during model fitting, regularized ML methods have been combined with filters using two-stage approaches. Unlike traditional two-stage GWAS,[155–157] these combined methods add an ML step to hypothesis testing using either data split or resampling. This combination attempts to balance sensitivity with false discovery by using screening tests to detect low signal, linear interactions and ML to train cross-validated prediction models of non-linear interactions.[143] In principle, this staged design can also take advantage of the plateau in statistical power curve at high sample sizes to contribute cases to an ML method where learning error often obeys an inverse polynomial function (**Figure 3**).[158–160]

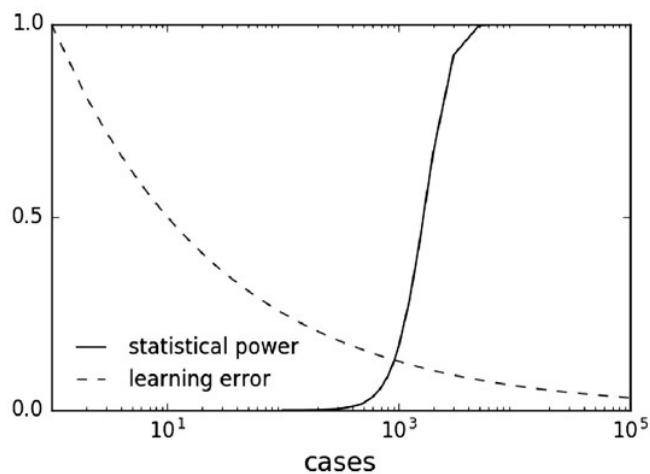

**Figure 3**: Sample plots of statistical power and learning curve error. Statistical power graph derived using Genomic Association Studies power calculator.[161] Learning curve assuming an inverse power law common to multiple machine learning methods.[158–160] Reproduced from Kang et al.[9] with permission by publisher Frontiers (https://www.frontiersin.org)

"Screen and clean" is an example of a two-stage model that performs lasso regression ("screen") on split data using a dictionary of pairwise interaction terms that is then followed by hypothesis testing ("clean") on remaining data.[134] This method has been applied to model SNP interaction and main effects.[162] A similar method reverses the order of lasso and hypothesis testing.[163]

The COMBI method for GWAS combines a linear support vector machine (with SNPs weighted using a moving average filter) and hypothesis testing using resampling to correct for familywise error rate.[164] Using training data from a 2007 Wellcome Trust Case Control Consortium GWAS study, Mieth et al. report higher power and precision to detect SNPs replicated in independent GWASes from 2008-2015 as compared to traditional and "screen and clean" based methods. COMBI also had fewer false discoveries, with 80% of discovered SNPs being validated in prior studies.

Similar two-stage approaches have been used with random forest (RF) and chi-squared test to select for SNPs.[165,166] RF and similar ensemble-tree model have been very popular due having built-in feature importance measures.[167] In simulations of GWAS, variable importance measures using RF demonstrated superior power to discover associations with disease compared to linear regression and dimensionality reduction methods.[168] While RF can provide measures of variable importance, there are limitations to inference and interpretability for variables. Generally, RFs and related approaches do not do variable selection—meaning, (a) there is no dimension reduction and (b) it is impossible to directly carry out tests of significance (without methods like resampling of synthetic data[168]), especially when the number of predictors approaches or exceeds the number of samples. Additionally—and perhaps as a result—there is disagreement about RF's ability to detect pure epistatic interactions compared to marginal effects.[169,170]

### 5.4.3 Unsupervised learning: feature transformation/construction

Feature transformations are popular methods for both decreasing the number of features and reconstructing the data to improve predictive performance. These methods compress the original features into fewer new ones that aim to preserve most of the original information. Unsupervised learning methods are used as they do not require labeled outputs and instead look for interesting substructures or relationships. These relationships can be between inputs or between inputs and outputs.

Clustering is a general class of methods that groups observed features into fewer features called centroids and assigns every data point to one of these centroids. The most popular clustering method for feature transformation is k-means clustering. Another popular method is hierarchical clustering, which aims to preserve the bottom-up tiered processes seen in biology.

Another popular transformation method is principal component analysis (PCA), which is motivated by eigenvector analysis. PCA finds the directions of highest variance in the dataset that are statistically uncorrelated (i.e., all orthogonal to each other). The direction of largest variance is the first principal component (PC), the second highest variance is the second PC, and so on. While PCA is often used for feature selection, it is also used in GWAS to control for potential confounding by ancestry or batch effects.[171,172]

Methods like k-means and PCA can be effective for finding basic substructures (well-separated, globular shapes) but do not work well for more irregular shapes. A class of methods called manifold learning algorithms is aimed to find low dimensional structure in high dimensional data. Multidimensional scaling (MDS) is one such method that aims to preserve information about the proximity of data points to each other. In genomics, MDS is used as an alternative to PCA.[150] A newer method called t-SNE finds a two-dimensional representation of data such that the distances between points in the 2D scatterplot match the distances between the same points in the original high dimensional dataset.[173–175]

Kernel-based methods such as support vector machines can transform features in higher dimensions where analysis may be easier.[176] Neural networks are another popular ML method that specializes in constructing features within the hidden layers after being initialized with observed features.

### 5.4.4 Multi-step feature selection in radiogenomics

Oh et al. apply several feature selection methods to GWAS data to create preconditioned random forest regression (PRFR) for prediction of rectal bleeding and erectile dysfunction after prostate radiotherapy.[88] The SNPs are initially pre-filtered for quality control, resulting in approximately 600,000 SNPSs in 368 patients. Stepwise forward selection of the SNPs into PCA was performed through inputting the first 1-2 principal components into logistic regression to predict binary toxicity outputs. By mapping transformed SNPs in logistic regression to predict binary toxicities, continuous "pre-conditioned outcomes" are created which are highly correlated with both the binary toxicities and the inputs.[177] The original SNPs were input into random forest regression to predict the preconditioned outcomes and validated on holdouts of the original samples. SNP importance rank was assessed using out-of-bag performance and inclusion in known biological pathways. Using this method, several genes implicated in ion transport and vascular regulation were found to be important for rectal bleeding and erectile dysfunction, respectively. This process was also applied to predict genitourinary toxicity and find associated gene ontology terms.[178]

**Table 3** summarizes PRFR and select radiogenomics models previously discussed.

**Table 3: Summary of select radiogenomic models that were discussed in this review.** SNP: single nucleotide polymorphism; UM: University of Michigan; LC: local control; RP2: Grade 2+ radiation pneumonitis; PORTOS: Post-Operative Radiation Therapy Outcomes Score[76]; DMFS: distant metastasis free survival; RSI/GARD: Radiosensitivity Index/Genome-Adjusted Radiation Dose[81]; PRFR: Preconditioned Random Forest Regression[88]; PCA: principal component analysis

| Radiogenomic model | Application | Model type | Feature selection method |
|---|---|---|---|
| Chi-square | genome-wide significance of SNPs | statistical inference | unvariate filter |
| UM Bayesian network | balance LC & RP2 after lung radiation | interpretable graph | stepwise wrapper |
| PORTOS | DMFS after prostatic-bed radiation | augmented statistical | candidate gene & filtering |
| RSI/GARD | various tumors' survival in vitro/in vivo | augmented mechanistic | uni-/multi-variable regression |
| PRFR | toxicity after prostate radiation | transformed random forest | filter, stepwise PCA |

# 6 Conclusion

As polygenetic assays to guide targeted systemic therapy becomes more established, the next logical step is to use genomics of tumor and normal tissues to better counsel patients and, ultimately, improve the therapeutic ratio in radiation oncology. As discussed, several tumor radiosensitivity panels have been developed and continue to undergo validation, including GARD, PORTOS and the Bayesian network approach for multi-omics. The PORTOS assay for post-prostatectomy radiation is now listed in the NCCN guidelines under the GenomeDX Decipher® platform.[25] PRFR and other radiotoxicity models based on SNPs identified via GWAS are also undergoing active development. The high-dimensionality and lower signal-to-noise in GWAS has been challenging, but efforts through the Radiogenomics Consortium to pool cohorts and apply new methodologies aim to address these issues.

New cooperative group phase III trials led by radiation oncology investigators have begun to incorporate genomic considerations in primary and secondary endpoints.[179–181] The future of genomics in radiation oncology is promising but will require multidisciplinary collaboration between physicians, radiobiologists, and medical physicists/informaticians in cooperative group settings to best utilize the spectrum of clinical applicability, biological considerations, and analytic principles.

# 7  Acknowledgements

The authors thank Ane Appelt for her suggestions and critique. This research has been supported by grants and contracts as follows: JK by the Radiological Society of North America Research & Education Foundation Resident Research Grant #RR1843; RLS by National Institutes of Mental Health 5R01MH075017; BSR from the U.S. National Institutes of Health (1R01CA134444 and HHSN261201500043C) and the U.S. Department of Defense (PC074201 and PC140371); SLK by National Institutes of Health/National Cancer Institute 1K07CA187546.

# 8  References

1. Burnet NG, Elliott RM, Dunning A, West CML. Radiosensitivity, radiogenomics and RAPPER. *Clin Oncol (R Coll Radiol)*. 2006;18(7):525-528.

2. West CM, Barnett GC. Genetics and genomics of radiotherapy toxicity: towards prediction. *Genome Med*. 2011;3(8):52. doi:10.1186/gm268

3. Story MD, Durante M. Radiogenomics. *Med Phys*. 2018;45(11):e1111-e1122. doi:10.1002/mp.13064

4. Hall WA, Bergom C, Thompson RF, et al. Precision Oncology and Genomically Guided Radiation Therapy: A Report From the American Society for Radiation Oncology/American Association of Physicists in Medicine/National Cancer Institute Precision Medicine Conference. *Int J Radiat Oncol Biol Phys*. June 2017. doi:10.1016/j.ijrobp.2017.05.044

5. Baumann M, Krause M, Overgaard J, et al. Radiation oncology in the era of precision medicine. *Nat Rev Cancer*. 2016;16(4):234-249. doi:10.1038/nrc.2016.18

6. Chen X, Ishwaran H. Random Forests for Genomic Data Analysis. *Genomics*. 2012;99(6):323-329. doi:10.1016/j.ygeno.2012.04.003

7. Libbrecht MW, Noble WS. Machine learning applications in genetics and genomics. *Nat Rev Genet*. 2015;16(6):321-332. doi:10.1038/nrg3920

8. Noble WS. What is a support vector machine? *Nature biotechnology*. 2006;24(12):1565-1567. doi:10.1038/nbt1206-1565

9. Kang J, Rancati T, Lee S, et al. Machine Learning and Radiogenomics: Lessons Learned and Future Directions. *Front Oncol*. 2018;8:228. doi:10.3389/fonc.2018.00228

10. Kang J, Schwartz R, Flickinger J, Beriwal S. Machine Learning Approaches for Predicting Radiation Therapy Outcomes: A Clinician's Perspective. *Int J Radiat Oncol Biol Phys*. 2015;93(5):1127-1135. doi:10.1016/j.ijrobp.2015.07.2286

11. El Naqa I, Li R, Murphy MJ. *Machine Learning in Radiation Oncology: Theory and Applications*. 1st ed. Springer International Publishing; 2015.

12. El Naqa I, Brock K, Yu Y, Langen K, Klein EE. On the Fuzziness of Machine Learning, Neural Networks, and Artificial Intelligence in Radiation Oncology. *Int J Radiat Oncol*. 2018;100(1):1-4. doi:10.1016/j.ijrobp.2017.06.011

13. Feng M, Valdes G, Dixit N, Solberg TD. Machine Learning in Radiation Oncology: Opportunities, Requirements, and Needs. *Front Oncol*. 2018;8. doi:10.3389/fonc.2018.00110

14. Thompson RF, Valdes G, Fuller CD, et al. Artificial intelligence in radiation oncology: A specialty-wide disruptive transformation? *Radiother Oncol*. 2018;129(3):421-426. doi:10.1016/j.radonc.2018.05.030

15. Mayo CS, Phillips M, McNutt TR, et al. Treatment data and technical process challenges for practical big data efforts in radiation oncology. *Medical Physics*. 2018;45(10):e793-e810. doi:10.1002/mp.13114

16. Prasad V, Gale RP. What precisely is precision oncology-and will it work.

17. Office of the Commissioner. Press Announcements - FDA approves first cancer treatment for any solid tumor with a specific genetic feature. https://www.fda.gov/NewsEvents/Newsroom/PressAnnouncements/ucm560167.htm. Accessed March 8, 2019.

18. Le DT, Uram JN, Wang H, et al. PD-1 Blockade in Tumors with Mismatch-Repair Deficiency. *New England Journal of Medicine*. 2015;372(26):2509-2520. doi:10.1056/NEJMoa1500596

19. NCI-sponsored trials in precision medicine | Major Initiatives | DCTD. https://dctd.cancer.gov/majorinitiatives/NCI-sponsored_trials_in_precision_medicine.htm. Accessed March 8, 2019.

20. Coyne GO, Takebe N, Chen AP. Defining precision: The precision medicine initiative trials NCI-MPACT and NCI-MATCH. *Curr Probl Cancer*. 2017;41(3):182-193. doi:10.1016/j.currproblcancer.2017.02.001

21. Clark GM. Prognostic factors versus predictive factors: Examples from a clinical trial of erlotinib. *Molecular Oncology*. 2008;1(4):406-412. doi:10.1016/j.molonc.2007.12.001

22. Ballman KV. Biomarker: Predictive or Prognostic? *J Clin Oncol*. 2015;33(33):3968-3971. doi:10.1200/JCO.2015.63.3651

23. Spratt DE. Ki-67 Remains Solely a Prognostic Biomarker in Localized Prostate Cancer - International Journal of Radiation Oncology • Biology • Physics. 2019. doi:doi:10.1016/j.ijrobp.2018.03.008

24. Tosoian JJ, Mamawala M, Epstein JI, et al. Intermediate and Longer-Term Outcomes From a Prospective Active-Surveillance Program for Favorable-Risk Prostate Cancer. *JCO*. 2015;33(30):3379-3385. doi:10.1200/JCO.2015.62.5764

25. National Comprehensive Cancer Network (NCCN) Clinical Practice Guidelines in Oncology. Prostate Cancer (Version 1.2019 - March 6, 2019). Prostate Cancer. https://www.nccn.org/professionals/physician_gls/PDF/prostate.pdf. Accessed March 15, 2019.

26. Paik S, Shak S, Tang G, et al. A Multigene Assay to Predict Recurrence of Tamoxifen-Treated, Node-Negative Breast Cancer. *New England Journal of Medicine*. 2004;351(27):2817-2826. doi:10.1056/NEJMoa041588

27. Sparano JA, Gray RJ, Makower DF, et al. Prospective Validation of a 21-Gene Expression Assay in Breast Cancer. *New England Journal of Medicine*. 2015;373(21):2005-2014. doi:10.1056/NEJMoa1510764

28. Speers C, Zhao S, Liu M, Bartelink H, Pierce LJ, Feng FY. Development and Validation of a Novel Radiosensitivity Signature in Human Breast Cancer. *Clin Cancer Res*. 2015;21(16):3667-3677. doi:10.1158/1078-0432.CCR-14-2898

29. Hegi ME, Stupp R. Withholding temozolomide in glioblastoma patients with unmethylated MGMT promoter—still a dilemma? *Neuro Oncol*. 2015;17(11):1425-1427. doi:10.1093/neuonc/nov198

30. Perry JR, Laperriere N, O'Callaghan CJ, et al. Short-Course Radiation plus Temozolomide in Elderly Patients with Glioblastoma. *http://dx.doi.org/101056/NEJMoa1611977*. March 2017. doi:NJ201703163761107

31. Hegi ME, Diserens AC, Gorlia T, et al. MGMT gene silencing and benefit from temozolomide in glioblastoma. *The New England journal of medicine*. 2005;352(10):997-1003. doi:10.1056/NEJMoa043331

32. Herrlinger U, Tzaridis T, Mack F, et al. Lomustine-temozolomide combination therapy versus standard temozolomide therapy in patients with newly diagnosed glioblastoma with methylated MGMT promoter (CeTeG/NOA–09): a randomised, open-label, phase 3 trial. *The Lancet*. 2019;393(10172):678-688. doi:10.1016/S0140-6736(18)31791-4

33. Spratt DE, Wahl DR, Lawrence TS. Genomic-adjusted radiation dose. *Lancet Oncol*. 2017;18(3):e127. doi:10.1016/s1470-2045(17)30092-x

34. Scott JG, Harrison LB, Torres-Roca JF. Genomic-adjusted radiation dose - Authors' reply. *Lancet Oncol*. 2017;18(3):e129. doi:10.1016/s1470-2045(17)30119-5

35. Italiano A. Prognostic or predictive? It's time to get back to definitions! *Journal of clinical oncology : official journal of the American Society of Clinical Oncology*. 2011;29(35):4718; author reply 4718-9. doi:10.1200/JCO.2011.38.3729

36. Amy R. Peck, Agnieszka K. Witkiewicz, Chengbao Liu, et al. Reply to A. Italiano. *Journal of Clinical Oncology*. 2011;29(35):4718-4719. doi:10.1200/jco.2011.38.5187

37. Matei D, Filiaci VL, Randall M, et al. A randomized phase III trial of cisplatin and tumor volume directed irradiation followed by carboplatin and paclitaxel vs. carboplatin and paclitaxel for optimally debulked, advanced endometrial carcinoma. *JCO*. 2017;35(15_suppl):5505-5505. doi:10.1200/JCO.2017.35.15_suppl.5505

38. Paz-Ares L, Luft A, Vicente D, et al. Pembrolizumab plus Chemotherapy for Squamous Non-Small-Cell Lung Cancer. *N Engl J Med*. 2018;379(21):2040-2051. doi:10.1056/NEJMoa1810865

39. Gandhi L, Rodríguez-Abreu D, Gadgeel S, et al. Pembrolizumab plus Chemotherapy in Metastatic Non–Small-Cell Lung Cancer. *New England Journal of Medicine*. April 2018. doi:10.1056/NEJMoa1801005

40. Bentzen SM, Constine LS, Deasy JO, et al. Quantitative Analyses of Normal Tissue Effects in the Clinic (QUANTEC): an introduction to the scientific issues. *Int J Radiat Oncol Biol Phys*. 2010;76(3 Suppl):S3-9. doi:10.1016/j.ijrobp.2009.09.040

41. Videtic GMM, Donington J, Giuliani M, et al. Stereotactic body radiation therapy for early-stage non-small cell lung cancer: Executive Summary of an ASTRO Evidence-Based Guideline. *Pract Radiat Oncol*. 2017;7(5):295-301. doi:10.1016/j.prro.2017.04.014

42. Morgan SC, Hoffman K, Loblaw DA, et al. Hypofractionated Radiation Therapy for Localized Prostate Cancer: Executive Summary of an ASTRO, ASCO, and AUA Evidence-Based Guideline. *Pract Radiat Oncol*. 2018;8(6):354-360. doi:10.1016/j.prro.2018.08.002

43. Smith BD, Bellon JR, Blitzblau R, et al. Radiation therapy for the whole breast: Executive summary of an American Society for Radiation Oncology (ASTRO) evidence-based guideline. *Pract Radiat Oncol*. 2018;8(3):145-152. doi:10.1016/j.prro.2018.01.012

44. Constine LS, Ronckers CM, Hua CH, et al. Pediatric Normal Tissue Effects in the Clinic (PENTEC): An International Collaboration to Analyse Normal Tissue Radiation Dose-Volume Response Relationships for Paediatric Cancer Patients. *Clinical oncology (Royal College of Radiologists (Great Britain))*. 2019;31(3):199-207. doi:10.1016/j.clon.2019.01.002

45. Milano MT, Grimm J, Soltys SG, et al. Single- and Multi-Fraction Stereotactic Radiosurgery Dose Tolerances of the Optic Pathways. *Int J Radiat Oncol Biol Phys*. January 2018. doi:10.1016/j.ijrobp.2018.01.053

46. Doroslovacki P, Tamhankar MA, Liu GT, Shindler KS, Ying GS, Alonso-Basanta M. Factors Associated with Occurrence of Radiation-induced Optic Neuropathy at "Safe" Radiation Dosage. *Seminars in ophthalmology*. 2018;33(4):581-588. doi:10.1080/08820538.2017.1346133

47. Bentzen SM, Overgaard M, Overgaard J. Clinical correlations between late normal tissue endpoints after radiotherapy: implications for predictive assays of radiosensitivity. *Eur J Cancer*. 1993;29A(10):1373-1376.

48. Tucker SL, Turesson I, Thames HD. Evidence for individual differences in the radiosensitivity of human skin. *Eur J Cancer*. 1992;28A(11):1783-1791.

49. Safwat A, Bentzen SM, Turesson I, Hendry JH. Deterministic rather than stochastic factors explain most of the variation in the expression of skin telangiectasia after radiotherapy. *Int J Radiat Oncol Biol Phys*. 2002;52(1):198-204.

50. Andreassen CN, Alsner J. Genetic variants and normal tissue toxicity after radiotherapy: a systematic review. *Radiother Oncol*. 2009;92(3):299-309. doi:10.1016/j.radonc.2009.06.015

51. Barnett GC, Coles CE, Elliott RM, et al. Independent validation of genes and polymorphisms reported to be associated with radiation toxicity: a prospective analysis study. *Lancet Oncol*. 2012;13(1):65-77. doi:10.1016/S1470-2045(11)70302-3

52. Andreassen CN, Schack LMH, Laursen LV, Alsner J. Radiogenomics - current status, challenges and future directions. *Cancer Lett*. 2016;382(1):127-136. doi:10.1016/j.canlet.2016.01.035

53. West C, Rosenstein BS. Establishment of a Radiogenomics Consortium. *International Journal of Radiation Oncology • Biology • Physics*. 2010;76(5):1295-1296. doi:10.1016/j.ijrobp.2009.12.017

54. West C, Rosenstein BS. Establishment of a radiogenomics consortium. *Radiother Oncol*. 2010;94(1):117-118. doi:10.1016/j.radonc.2009.12.007

55. Talbot CJ, Tanteles GA, Barnett GC, et al. A replicated association between polymorphisms near TNFα and risk for adverse reactions to radiotherapy. *Br J Cancer*. 2012;107(4):748-753. doi:10.1038/bjc.2012.290

56. Seibold P, Behrens S, Schmezer P, et al. XRCC1 Polymorphism Associated With Late Toxicity After Radiation Therapy in Breast Cancer Patients. *Int J Radiat Oncol Biol Phys*. 2015;92(5):1084-1092. doi:10.1016/j.ijrobp.2015.04.011

57. Andreassen CN, Rosenstein BS, Kerns SL, et al. Individual patient data meta-analysis shows a significant association between the ATM rs1801516 SNP and toxicity after radiotherapy in 5456 breast and prostate cancer patients. *Radiother Oncol*. 2016;121(3):431-439. doi:10.1016/j.radonc.2016.06.017

58. Pang Q, Wei Q, Xu T, et al. Functional promoter variant rs2868371 of HSPB1 is associated with risk of radiation pneumonitis after chemoradiation for non-small cell lung cancer. *Int J Radiat Oncol Biol Phys*. 2013;85(5):1332-1339. doi:10.1016/j.ijrobp.2012.10.011

59. Lopez Guerra JL, Wei Q, Yuan X, et al. Functional promoter rs2868371 variant of HSPB1 associates with radiation-induced esophageal toxicity in patients with non-small-cell lung cancer treated with radio(chemo)therapy. *Radiother Oncol*. 2011;101(2):271-277. doi:10.1016/j.radonc.2011.08.039

60. Guerra JLL, Gomez D, Wei Q, et al. Association between single nucleotide polymorphisms of the transforming growth factor β1 gene and the risk of severe radiation esophagitis in patients with lung cancer. *Radiother Oncol*. 2012;105(3):299-304. doi:10.1016/j.radonc.2012.08.014

61. Barnett GC, Elliott RM, Alsner J, et al. Individual patient data meta-analysis shows no association between the SNP rs1800469 in TGFB and late radiotherapy toxicity. *Radiother Oncol*. 2012;105(3):289-295. doi:10.1016/j.radonc.2012.10.017

62. Grossberg AJ, Lei X, Xu T, et al. Association of Transforming Growth Factor β Polymorphism C-509T With Radiation-Induced Fibrosis Among Patients With Early-Stage Breast Cancer: A Secondary Analysis of a Randomized Clinical Trial. *JAMA Oncol*. 2018;4(12):1751-1757. doi:10.1001/jamaoncol.2018.2583

63. Rosenstein BS. Radiogenomics: Identification of Genomic Predictors for Radiation Toxicity. *Semin Radiat Oncol*. 2017;27(4):300-309. doi:10.1016/j.semradonc.2017.04.005

64. Fachal L, Gómez-Caamaño A, Barnett GC, et al. A three-stage genome-wide association study identifies a susceptibility locus for late radiotherapy toxicity at 2q24.1. *Nature Genetics*. 2014;46(8):891-894. doi:10.1038/ng.3020

65. Kerns SL, Dorling L, Fachal L, et al. Meta-analysis of Genome Wide Association Studies Identifies Genetic Markers of Late Toxicity Following Radiotherapy for Prostate Cancer. *EBioMedicine*. 2016;10:150-163. doi:10.1016/j.ebiom.2016.07.022

66. Zarrei M, MacDonald JR, Merico D, Scherer SW. A copy number variation map of the human genome. *Nature Reviews Genetics*. 2015;16(3):172-183. doi:10.1038/nrg3871

67. Pirooznia M, Goes FS, Zandi PP. Whole-genome CNV analysis: advances in computational approaches. *Front Genet*. 2015;6. doi:10.3389/fgene.2015.00138

68. Coates J, Jeyaseelan AK, Ybarra N, et al. Contrasting analytical and data-driven frameworks for radiogenomic modeling of normal tissue toxicities in prostate cancer. *Radiother Oncol*. 2015;115(1):107-113. doi:10.1016/j.radonc.2015.03.005

69. Yard BD, Adams DJ, Chie EK, et al. A genetic basis for the variation in the vulnerability of cancer to DNA damage. *Nat Commun*. 2016;7:11428. doi:10.1038/ncomms11428

70. El Naqa I, Kerns SL, Coates J, et al. Radiogenomics and radiotherapy response modeling. *Phys Med Biol*. 2017;62(16):R179-R206. doi:10.1088/1361-6560/aa7c55

71. Oh JH, Craft J, Al Lozi R, et al. A Bayesian network approach for modeling local failure in lung cancer. *Phys Med Biol*. 2011;56(6):1635-1651. doi:10.1088/0031-9155/56/6/008

72. Choudhury A, Nelson LD, Teo MTW, et al. MRE11 expression is predictive of cause-specific survival following radical radiotherapy for muscle-invasive bladder cancer. *Cancer Res*. 2010;70(18):7017-7026. doi:10.1158/0008-5472.CAN-10-1202

73. Laurberg JR, Brems-Eskildsen AS, Nordentoft I, et al. Expression of TIP60 (tat-interactive protein) and MRE11 (meiotic recombination 11 homolog) predict treatment-specific outcome of localised invasive bladder cancer. *BJU Int*. 2012;110(11 Pt C):E1228-1236. doi:10.1111/j.1464-410X.2012.11564.x

74. Teo MTW, Dyrskjøt L, Nsengimana J, et al. Next-generation sequencing identifies germline MRE11A variants as markers of radiotherapy outcomes in muscle-invasive bladder cancer. *Ann Oncol*. 2014;25(4):877-883. doi:10.1093/annonc/mdu014

75. Walker AK, Karaszi K, Valentine H, et al. MRE11 as a predictive biomarker of outcome following radiotherapy in bladder cancer. *International Journal of Radiation Oncology*Biology*Physics*. March 2019. doi:10.1016/j.ijrobp.2019.03.015

76. Zhao SG, Chang SL, Spratt DE, et al. Development and validation of a 24-gene predictor of response to postoperative radiotherapy in prostate cancer: a matched, retrospective analysis. *Lancet Oncol*. 2016;17(11):1612-1620. doi:10.1016/S1470-2045(16)30491-0

77. Torres-Roca JF. A molecular assay of tumor radiosensitivity: a roadmap towards biology-based personalized radiation therapy. *Per Med*. 2012;9(5):547-557. doi:10.2217/pme.12.55

78. Eschrich SA, Fulp WJ, Pawitan Y, et al. Validation of a radiosensitivity molecular signature in breast cancer. *Clinical cancer research : an official journal of the American Association for Cancer Research*. 2012;18(18):5134-5143. doi:10.1158/1078-0432.ccr-12-0891

79. Creelan B, Eschrich SA, Fulp WJ, Torres-Roca JF. A Gene Expression Platform to Predict Benefit From Adjuvant External Beam Radiation in Resected Non-Small Cell Lung Cancer. *International Journal of Radiation Oncology\*Biology\*Physics*. 2014;90(1, Supplement):S76-S77. doi:https://doi.org/10.1016/j.ijrobp.2014.05.455

80. Torres-Roca JF, Erho N, Vergara I, et al. A Molecular Signature of Radiosensitivity (RSI) is an RT-specific Biomarker in Prostate Cancer. *International Journal of Radiation Oncology\*Biology\*Physics*. 2014;90(1, Supplement):S157. doi:https://doi.org/10.1016/j.ijrobp.2014.05.642

81. Scott JG, Berglund A, Schell MJ, et al. A genome-based model for adjusting radiotherapy dose (GARD): a retrospective, cohort-based study. *Lancet Oncol*. 2017;18(2):202-211. doi:10.1016/S1470-2045(16)30648-9

82. Moertl S, Mutschelknaus L, Heider T, Atkinson MJ. MicroRNAs as novel elements in personalized radiotherapy. *Translational Cancer Research*. 2016;5(6):S1262-S1269-S1269. doi:10.21037/10576

83. Sun Y, Hawkins PG, Bi N, et al. Serum MicroRNA Signature Predicts Response to High-Dose Radiation Therapy in Locally Advanced Non-Small Cell Lung Cancer. *Int J Radiat Oncol Biol Phys*. 2018;100(1):107-114. doi:10.1016/j.ijrobp.2017.08.039

84. Summerer I, Niyazi M, Unger K, et al. Changes in circulating microRNAs after radiochemotherapy in head and neck cancer patients. *Radiation Oncology*. 2013;8(1):296. doi:10.1186/1748-717X-8-296

85. Higgins KA, Saba NF, Shin DM, et al. Circulating Pre-treatment miRNAs as Potential Biomarkers to Predict Radiation Toxicity. *International Journal of Radiation Oncology • Biology • Physics*. 2017;99(2):E596. doi:10.1016/j.ijrobp.2017.06.2035

86. Templin T, Paul S, Amundson SA, et al. RADIATION-INDUCED MICRO-RNA EXPRESSION CHANGES IN PERIPHERAL BLOOD CELLS OF RADIOTHERAPY PATIENTS. *Int J Radiat Oncol Biol Phys*. 2011;80(2):549-557. doi:10.1016/j.ijrobp.2010.12.061

87. Merchant TE, Bendel AE, Sabin ND, et al. Conformal Radiation Therapy for Pediatric Ependymoma, Chemotherapy for Incompletely Resected Ependymoma, and Observation for Completely Resected, Supratentorial Ependymoma. *JCO*. February 2019:JCO.18.01765. doi:10.1200/JCO.18.01765

88. Oh JH, Kerns S, Ostrer H, Powell SN, Rosenstein B, Deasy JO. Computational methods using genome-wide association studies to predict radiotherapy complications and to identify correlative molecular processes. *Sci Rep*. 2017;7:43381. doi:10.1038/srep43381

89. Trott K-R, Doerr W, Facoetti A, et al. Biological mechanisms of normal tissue damage: importance for the design of NTCP models. *Radiother Oncol*. 2012;105(1):79-85. doi:10.1016/j.radonc.2012.05.008

90. Onjukka E, Baker C, Nahum A. The performance of normal-tissue complication probability models in the presence of confounding factors. *Med Phys*. 2015;42(5):2326-2341. doi:10.1118/1.4917219

91. Kutcher GJ, Burman C. Calculation of complication probability factors for non-uniform normal tissue irradiation: the effective volume method. *Int J Radiat Oncol Biol Phys*. 1989;16(6):1623-1630.

92. Coates J. Motivation for the inclusion of genetic risk factors of radiosensitivity alongside dosimetric and clinical parameters in predicting normal tissue effects. *Acta Oncologica*. 2015;54(8):1230-1231. doi:10.3109/0284186X.2014.999163

93. Tucker SL, Li M, Xu T, et al. Incorporating single-nucleotide polymorphisms into the Lyman model to improve prediction of radiation pneumonitis. *Int J Radiat Oncol Biol Phys*. 2013;85(1):251-257. doi:10.1016/j.ijrobp.2012.02.021

94. Peeters STH, Hoogeman MS, Heemsbergen WD, Hart AAM, Koper PCM, Lebesque JV. Rectal bleeding, fecal incontinence, and high stool frequency after conformal radiotherapy for prostate cancer: normal tissue complication probability modeling. *Int J Radiat Oncol Biol Phys*. 2006;66(1):11-19. doi:10.1016/j.ijrobp.2006.03.034

95. Kang J, Steward RL, Kim Y, Schwartz RS, LeDuc PR, Puskar KM. Response of an actin filament network model under cyclic stretching through a coarse grained Monte Carlo approach. *J Theor Biol*. 2011;274(1):109-119. doi:10.1016/j.jtbi.2011.01.011

96. Kang J, Puskar KM, Ehrlicher AJ, LeDuc PR, Schwartz RS. Structurally Governed Cell Mechanotransduction through Multiscale Modeling. *Scientific Reports*. 2015;5:8622. doi:10.1038/srep08622

97. Hope AJ, Lindsay PE, El Naqa I, et al. Modeling radiation pneumonitis risk with clinical, dosimetric, and spatial parameters. *Int J Radiat Oncol Biol Phys*. 2006;65(1):112-124. doi:10.1016/j.ijrobp.2005.11.046

98. Mansur YA, Rojano E, Ranea JAG, Perkins JR. Chapter 7 - Analyzing the Effects of Genetic Variation in Noncoding Genomic Regions. In: Deigner H-P, Kohl M, eds. *Precision Medicine*. Academic Press; 2018:119-144. doi:10.1016/B978-0-12-805364-5.00007-X

99. Guo Z, Shu Y, Zhou H, Zhang W, Wang H. Radiogenomics helps to achieve personalized therapy by evaluating patient responses to radiation treatment. *Carcinogenesis*. 2015;36(3):307-317. doi:10.1093/carcin/bgv007

100. Defraene G, Van den Bergh L, Al-Mamgani A, et al. The benefits of including clinical factors in rectal normal tissue complication probability modeling after radiotherapy for prostate cancer. *Int J Radiat Oncol Biol Phys*. 2012;82(3):1233-1242. doi:10.1016/j.ijrobp.2011.03.056

101. Emami B, Lyman J, Brown A, et al. Tolerance of normal tissue to therapeutic irradiation. *Int J Radiat Oncol Biol Phys*. 1991;21(1):109-122.

102. Tucker SL, Liu HH, Liao Z, et al. Analysis of radiation pneumonitis risk using a generalized Lyman model. *Int J Radiat Oncol Biol Phys*. 2008;72(2):568-574. doi:10.1016/j.ijrobp.2008.04.053

103. Abler D, Kanellopoulos V, Davies J, et al. Data-driven Markov models and their application in the evaluation of adverse events in radiotherapy. *J Radiat Res*. 2013;54(suppl_1):i49-i55. doi:10.1093/jrr/rrt040

104.    El Naqa I, Suneja G, Lindsay PE, et al. Dose response explorer: an integrated open-source tool for exploring and modelling radiotherapy dose-volume outcome relationships. *Phys Med Biol*. 2006;51(22):5719-5735. doi:10.1088/0031-9155/51/22/001

105.    Breiman L. Statistical Modeling: The Two Cultures (with comments and a rejoinder by the author). *Statist Sci*. 2001;16(3):199-231. doi:10.1214/ss/1009213726

106.    Valdes G, Luna JM, Eaton E, Simone CB, Ungar LH, Solberg TD. MediBoost: a Patient Stratification Tool for Interpretable Decision Making in the Era of Precision Medicine. *Sci Rep*. 2016;6:37854. doi:10.1038/srep37854

107.    Caruana R, Lou Y, Gehrke J, Koch P, Sturm M, Elhadad N. Intelligible Models for HealthCare: Predicting Pneumonia Risk and Hospital 30-day Readmission. In: *Proceedings of the 21th ACM SIGKDD International Conference on Knowledge Discovery and Data Mining*. KDD '15. New York, NY, USA: ACM; 2015:1721–1730. doi:10.1145/2783258.2788613

108.    FDA, U. S. Food & Drug Administration. *Clinical and Patient Decision Support Software Draft Guidance*.; 2017.

109.    Ghorbani A, Abid A, Zou J. Interpretation of Neural Networks is Fragile. *arXiv:171010547 [cs, stat]*. October 2017. http://arxiv.org/abs/1710.10547. Accessed June 20, 2019.

110.    Interpretable ML Symposium - NIPS 2017. http://interpretable.ml/. Accessed April 7, 2019.

111.    Christopher M. Bishop. *Pattern Recognition and Machine Learning (Information Science and Statistics)*. Springer-Verlag New York, Inc.; 2006.

112.    Wasserman L. *All of Statistics: A Concise Course in Statistical Inference*. New York: Springer-Verlag; 2004. https://www.springer.com/us/book/9780387402727. Accessed March 11, 2019.

113.    Chen Y-C, Wheeler TA, Kochenderfer MJ. Learning Discrete Bayesian Networks from Continuous Data. *arXiv:151202406 [cs]*. December 2015. http://arxiv.org/abs/1512.02406. Accessed March 11, 2019.

114.    Lee S, Ybarra N, Jeyaseelan K, et al. Bayesian network ensemble as a multivariate strategy to predict radiation pneumonitis risk. *Med Phys*. 2015;42(5):2421-2430. doi:10.1118/1.4915284

115.    Luo Y, El Naqa I, McShan DL, et al. Unraveling biophysical interactions of radiation pneumonitis in non-small-cell lung cancer via Bayesian network analysis. *Radiother Oncol*. 2017;123(1):85-92. doi:10.1016/j.radonc.2017.02.004

116.    Aliferis CF, Tsamardinos I, Statnikov A. HITON: A Novel Markov Blanket Algorithm for Optimal Variable Selection. *AMIA Annu Symp Proc*. 2003;2003:21-25.

117.    Fu S, Desmarais MC. Markov Blanket based Feature Selection: A Review of Past Decade. 2010:9.

118.    Luo Y, McShan DL, Matuszak MM, et al. A multiobjective Bayesian networks approach for joint prediction of tumor local control and radiation pneumonitis in nonsmall-cell lung cancer (NSCLC) for response-adapted radiotherapy. *Med Phys*. June 2018. doi:10.1002/mp.13029

119.     Hall EJ, Giaccia AJ. *Radiobiology for the Radiologist*. Philadelphia: Wolters Kluwer Health/Lippincott Williams & Wilkins; 2012.

120.     Boyd S, Vandenberghe L. *Convex Optimization*. New York, NY, USA: Cambridge University Press; 2004.

121.     Wong Y-N, Egleston BL, Sachdeva K, et al. Cancer patients' trade-offs among efficacy, toxicity and out-of-pocket cost in the curative and non-curative setting. *Med Care*. 2013;51(9). doi:10.1097/MLR.0b013e31829faffd

122.     Loh KP, Mohile SG, Epstein RM, et al. Willingness to bear adversity and beliefs about the curability of advanced cancer. *JCO*. 2018;36(34_suppl):20-20. doi:10.1200/JCO.2018.36.34_suppl.20

123.     Tischer B, Bilang M, Kraemer M, Ronga P, Lacouture ME. A survey of patient and physician acceptance of skin toxicities from anti-epidermal growth factor receptor therapies. *Support Care Cancer*. 2018;26(4):1169-1179. doi:10.1007/s00520-017-3938-7

124.     Di Maio M, Gallo C, Leighl NB, et al. Symptomatic Toxicities Experienced During Anticancer Treatment: Agreement Between Patient and Physician Reporting in Three Randomized Trials. *JCO*. 2015;33(8):910-915. doi:10.1200/JCO.2014.57.9334

125.     Di Maio M, Basch E, Bryce J, Perrone F. Patient-reported outcomes in the evaluation of toxicity of anticancer treatments. *Nat Rev Clin Oncol*. 2016;13(5):319-325. doi:10.1038/nrclinonc.2015.222

126.     Moore GE. Cramming more components onto integrated circuits, Reprinted from Electronics, volume 38, number 8, April 19, 1965, pp.114 ff. *IEEE Solid-State Circuits Society Newsletter*. 2006;11(3):33-35. doi:10.1109/N-SSC.2006.4785860

127.     Uppu S, Krishna A, Gopalan RP. A Review on Methods for Detecting SNP Interactions in High-Dimensional Genomic Data. *IEEE/ACM Transactions on Computational Biology and Bioinformatics*. 2018;15(2):599-612. doi:10.1109/TCBB.2016.2635125

128.     Efroymson MA. Multiple regression analysis. In: *Mathematical Methods for Digital Computers*. New York, NY: Wiley; 1960:191-203.

129.     Harrell F. *Regression Modeling Strategies: With Applications to Linear Models, Logistic Regression, and Survival Analysis*. New York: Springer-Verlag; 2001. https://www.springer.com/us/book/9781441929181. Accessed April 3, 2019.

130.     Fan J, Lv J. Sure independence screening for ultrahigh dimensional feature space. *Journal of the Royal Statistical Society: Series B (Statistical Methodology)*. 2008;70(5):849-911. doi:10.1111/j.1467-9868.2008.00674.x

131.     Fan J, Samworth R, Wu Y. Ultrahigh Dimensional Feature Selection: Beyond The Linear Model. *Journal of Machine Learning Research*. 2009;10(Sep):2013-2038.

132.     Tibshirani R. Regression Shrinkage and Selection via the Lasso. *Journal of the Royal Statistical Society Series B (Methodological)*. 1996;58(1):267-288.

133.    Zou H, Hastie T. Regularization and variable selection via the elastic net. *J Roy Stat Soc Ser B (Stat Method)*. 2005;67(2):301-320. doi:10.1111/j.1467-9868.2005.00503.x

134.    Wasserman L, Roeder K. High Dimensional Variable Selection. *Ann Stat*. 2009;37(5A):2178-2201.

135.    Isabelle Guyon, André Elisseeff. An introduction to variable and feature selection. *J Mach Learn Res*. 2003;3:1157-1182.

136.    Pongpanich M, Sullivan PF, Tzeng J-Y. A quality control algorithm for filtering SNPs in genome-wide association studies. *Bioinformatics*. 2010;26(14):1731-1737. doi:10.1093/bioinformatics/btq272

137.    Turner S, Armstrong LL, Bradford Y, et al. Quality Control Procedures for Genome Wide Association Studies. *Curr Protoc Hum Genet*. 2011;CHAPTER:Unit1.19. doi:10.1002/0471142905.hg0119s68

138.    Sterne JAC, Smith GD. Sifting the evidence—what's wrong with significance tests? *BMJ*. 2001;322(7280):226-231.

139.    Johnson RC, Nelson GW, Troyer JL, et al. Accounting for multiple comparisons in a genome-wide association study (GWAS). *BMC genomics*. 2010;11:724. doi:10.1186/1471-2164-11-724

140.    Benjamini Y, Hochberg Y. Controlling the False Discovery Rate: A Practical and Powerful Approach to Multiple Testing. *Journal of the Royal Statistical Society Series B (Methodological)*. 1995;57(1):289-300.

141.    Kerns SL, Ostrer H, Stock R, et al. Genome Wide Association Study to Identify Single Nucleotide Polymorphisms (SNPs) Associated with the Development of Erectile Dysfunction in African-American Men Following Radiotherapy for Prostate Cancer. *Int J Radiat Oncol Biol Phys*. 2010;78(5):1292-1300. doi:10.1016/j.ijrobp.2010.07.036

142.    Cordell HJ. Epistasis: what it means, what it doesn't mean, and statistical methods to detect it in humans. *Hum Mol Genet*. 2002;11(20):2463-2468.

143.    Wei WH, Hemani G, Haley CS. Detecting epistasis in human complex traits. *Nature reviews Genetics*. 2014;15(11):722-733. doi:10.1038/nrg3747

144.    Fish AE, Capra JA, Bush WS. Are Interactions between cis-Regulatory Variants Evidence for Biological Epistasis or Statistical Artifacts? *Am J Hum Genet*. 2016;99(4):817-830. doi:10.1016/j.ajhg.2016.07.022

145.    Yang P, Ho JW, Yang YH, Zhou BB. Gene-gene interaction filtering with ensemble of filters. *BMC Bioinformatics*. 2011;12 Suppl 1:S10. doi:10.1186/1471-2105-12-S1-S10

146.    Moore JH. Epistasis analysis using ReliefF. *Methods Mol Biol*. 2015;1253:315-325. doi:10.1007/978-1-4939-2155-3_17

147.    Greene CS, Penrod NM, Kiralis J, Moore JH. Spatially uniform relieff (SURF) for computationally-efficient filtering of gene-gene interactions. *BioData Min*. 2009;2(1):5. doi:10.1186/1756-0381-2-5

148.     Kohavi R, John GH. Wrappers for feature subset selection. *Artif Intell*. 1997;97(1–2):273-324. doi:http://dx.doi.org/10.1016/S0004-3702(97)00043-X

149.     Coates J, Souhami L, El Naqa I. Big Data Analytics for Prostate Radiotherapy. *Front Oncol*. 2016;6:149. doi:10.3389/fonc.2016.00149

150.     Okser S, Pahikkala T, Airola A, Salakoski T, Ripatti S, Aittokallio T. Regularized machine learning in the genetic prediction of complex traits. *PLoS Genet*. 2014;10(11):e1004754. doi:10.1371/journal.pgen.1004754

151.     Hoerl AE, Kennard RW. Ridge regression: Biased estimation for nonorthogonal problems. *Technometrics*. 1970;12(1):55-67.

152.     Ogutu JO, Schulz-Streeck T, Piepho H-P. Genomic selection using regularized linear regression models: ridge regression, lasso, elastic net and their extensions. *BMC Proc*. 2012;6 Suppl 2:S10. doi:10.1186/1753-6561-6-S2-S10

153.     Wu TT, Chen YF, Hastie T, Sobel E, Lange K. Genome-wide association analysis by lasso penalized logistic regression. *Bioinformatics*. 2009;25(6):714-721. doi:10.1093/bioinformatics/btp041

154.     Meinshausen N, Bühlmann P. Stability selection. *Journal of the Royal Statistical Society: Series B (Statistical Methodology)*. 2010;72(4):417-473.

155.     Satagopan JM, Verbel DA, Venkatraman ES, Offit KE, Begg CB. Two-stage designs for gene-disease association studies. *Biometrics*. 2002;58(1):163-170.

156.     Skol AD, Scott LJ, Abecasis GR, Boehnke M. Joint analysis is more efficient than replication-based analysis for two-stage genome-wide association studies. *Nat Genet*. 2006;38(2):209-213. doi:10.1038/ng1706

157.     Skol AD, Scott LJ, Abecasis GR, Boehnke M. Optimal designs for two-stage genome-wide association studies. *Genet Epidemiol*. 2007;31(7):776-788. doi:10.1002/gepi.20240

158.     Mukherjee S, Tamayo P, Rogers S, et al. Estimating dataset size requirements for classifying DNA microarray data. *J Comput Biol*. 2003;10(2):119-142. doi:10.1089/106652703321825928

159.     Cortes C, Jackel LD, Solla SA, Vapnik V, Denker JS. Learning curves: Asymptotic values and rate of convergence. In: ; 1994:327-334.

160.     Dietrich R, Opper M, Sompolinsky H. Statistical Mechanics of Support Vector Networks. *Physical Review Letters*. 1999;82(14):2975-2978. doi:10.1103/PhysRevLett.82.2975

161.     Johnson JL, Abecasis GR. GAS Power Calculator: web-based power calculator for genetic association studies. *bioRxiv*. July 2017:164343. doi:10.1101/164343

162.     Wu J, Devlin B, Ringquist S, Trucco M, Roeder K. Screen and clean: a tool for identifying interactions in genome-wide association studies. *Genet Epidemiol*. 2010;34(3):275-285. doi:10.1002/gepi.20459

163.    Shi G, Boerwinkle E, Morrison AC, Gu CC, Chakravarti A, Rao DC. Mining gold dust under the genome wide significance level: a two-stage approach to analysis of GWAS. *Genet Epidemiol*. 2011;35(2):111-118. doi:10.1002/gepi.20556

164.    Mieth B, Kloft M, Rodriguez JA, et al. Combining Multiple Hypothesis Testing with Machine Learning Increases the Statistical Power of Genome-wide Association Studies. *Sci Rep*. 2016;6:36671. doi:10.1038/srep36671

165.    Nguyen TT, Huang J, Wu Q, Nguyen T, Li M. Genome-wide association data classification and SNPs selection using two-stage quality-based Random Forests. *BMC genomics*. 2015;16 Suppl 2:S5. doi:10.1186/1471-2164-16-S2-S5

166.    Roshan U, Chikkagoudar S, Wei Z, Wang K, Hakonarson H. Ranking causal variants and associated regions in genome-wide association studies by the support vector machine and random forest. *Nucleic Acids Res*. 2011;39(9):e62. doi:10.1093/nar/gkr064

167.    Lunetta KL, Hayward LB, Segal J, Van Eerdewegh P. Screening large-scale association study data: exploiting interactions using random forests. *BMC genetics*. 2004;5:32. doi:10.1186/1471-2156-5-32

168.    Molinaro AM, Carriero N, Bjornson R, Hartge P, Rothman N, Chatterjee N. Power of data mining methods to detect genetic associations and interactions. *Hum Hered*. 2011;72(2):85-97. doi:10.1159/000330579

169.    Yoshida M, Koike A. SNPInterForest: a new method for detecting epistatic interactions. *BMC Bioinformatics*. 2011;12:469. doi:10.1186/1471-2105-12-469

170.    Winham SJ, Colby CL, Freimuth RR, et al. SNP interaction detection with Random Forests in high-dimensional genetic data. *BMC Bioinformatics*. 2012;13:164. doi:10.1186/1471-2105-13-164

171.    Patterson N, Price AL, Reich D. Population structure and eigenanalysis. *PLoS Genet*. 2006;2(12):e190. doi:10.1371/journal.pgen.0020190

172.    Price AL, Patterson NJ, Plenge RM, Weinblatt ME, Shadick NA, Reich D. Principal components analysis corrects for stratification in genome-wide association studies. *Nat Genet*. 2006;38(8):904-909. doi:10.1038/ng1847

173.    Maaten L van der, Hinton G. Visualizing Data using t-SNE. *Journal of Machine Learning Research*. 2008;9(Nov):2579-2605.

174.    Wattenberg M, Viégas F, Johnson I. How to Use t-SNE Effectively. *Distill*. 2016;1(10):e2. doi:10.23915/distill.00002

175.    Li W, Cerise JE, Yang Y, Han H. Application of t-SNE to human genetic data. *J Bioinform Comput Biol*. 2017;15(04):1750017. doi:10.1142/S0219720017500172

176.    Cho BH, Yu H, Lee J, Chee YJ, Kim IY, Kim SI. Nonlinear support vector machine visualization for risk factor analysis using nomograms and localized radial basis function kernels. *IEEE Trans Inf Technol Biomed*. 2008;12(2):247-256. doi:10.1109/TITB.2007.902300

177.    Paul D, Bair E, Hastie T, Tibshirani R. "Preconditioning" for Feature Selection and Regression in High-Dimensional Problems. *The Annals of Statistics*. 2008;36(4):1595-1618.

178.    Lee S, Kerns S, Ostrer H, Rosenstein B, Deasy JO, Oh JH. Machine Learning on a Genome-wide Association Study to Predict Late Genitourinary Toxicity After Prostate Radiation Therapy. *Int J Radiat Oncol Biol Phys*. 2018;101(1):128-135. doi:10.1016/j.ijrobp.2018.01.054

179.    PORTEC-4a: Molecular Profile-based Versus Standard Adjuvant Radiotherapy in Endometrial Cancer - Full Text View - ClinicalTrials.gov. https://clinicaltrials.gov/ct2/show/NCT03469674. Accessed June 21, 2019.

180.    Wortman BG, Bosse T, Nout RA, et al. Molecular-integrated risk profile to determine adjuvant radiotherapy in endometrial cancer: Evaluation of the pilot phase of the PORTEC-4a trial. *Gynecol Oncol*. 2018;151(1):69-75. doi:10.1016/j.ygyno.2018.07.020

181.    Radiation Therapy With or Without Apalutamide in Treating Patients With Stage III-IV Prostate Cancer - Full Text View - ClinicalTrials.gov. https://clinicaltrials.gov/ct2/show/NCT03371719. Accessed June 21, 2019.